\documentstyle[12pt]{article}
\def\thebibliography#1{\bigskip\section*{\centering
References\\}\bigskip\list
{\arabic{enumi}.}{\settowidth\labelwidth{#1}\leftmargin\labelwidth
\advance\leftmargin\labelsep\usecounter{enumi}}
\def\newblock{\hskip .11em plus .33em minus .07em}
\sloppy\clubpenalty4000\widowpenalty4000 \sfcode`\.=1000\relax}

\def\op#1{\mathop{\fam0 #1}\limits}
\newcommand{\Id}{{\rm Id\,}}
\def\Ker{{\rm Ker\,}}
\def\Im{{\rm Im\,}}
\newcommand{\ben}{\begin{eqnarray}}
\newcommand{\een}{\end{eqnarray}}
\newcommand{\be}{\begin{eqnarray*}}
\newcommand{\ee}{\end{eqnarray*}}
\newcommand{\bea}{\begin{eqalph}}
\newcommand{\eea}{\end{eqalph}}
\newcommand{\cL}{{\cal L}}
\newcommand{\cE}{{\cal E}}
\newcommand{\cH}{{\cal H}}
\newcommand{\cF}{{\cal F}}
\newcommand{\cT}{{\cal T}}

\newcommand{\al}{\alpha}
\newcommand{\bt}{\beta}
\newcommand{\th}{\theta}
\newcommand{\Om}{\Omega}

\newcommand{\dl}{\delta}
\newcommand{\ap}{\approx}
\newcommand{\la}{\lambda}
\newcommand{\La}{\Lambda}
\newcommand{\om}{\omega}
\newcommand{\m}{\mu}
\newcommand{\n}{\nu}
\newcommand{\ot}{\otimes}
\newcommand{\kp}{\kappa}
\newcommand{\g}{\gamma}
\newcommand{\G}{\Gamma}
\newcommand{\ve}{\varepsilon}
\newcommand{\si}{\sigma}
\newcommand{\Si}{\Sigma}
\newcommand{\w}{\wedge}
\newcommand{\wt}{\widetilde}
\newcommand{\wh}{\widehat}
\newcommand{\ol}{\overline}
\newcommand{\dr}{\partial}
\newcounter{eqalph}
\newcounter{equationa}

\newenvironment{eqalph}{\stepcounter{equation}
\setcounter{equationa}{\value{equation}}
\setcounter{equation}{0}

\begin{eqnarray}}{\end{eqnarray}
\setcounter{equation}{\value{equationa}}}

\newenvironment{lemma}{{\sc Lemma.}}{$\Box$\medskip}

\begin{document}
\hbox{}

\centerline{\large\bf STRESS-ENERGY-MOMENTUM TENSORS}
\medskip

\centerline{\large\bf IN CONSTRAINT FIELD THEORIES}
\bigskip

\centerline{\bf Gennadi A. Sardanashvily}
\medskip

\centerline{Department of Theoretical Physics, Physics Faculty,}

\centerline{Moscow State University, 117234 Moscow, Russia}

\centerline{E-mail: sard@grav.phys.msu.su}
\bigskip

\begin{abstract}
One has not any conventional energy-momentum conservation law
in Lagrangian field theory, but relations involving different
stress-energy-momentum tensors associated with different connections.
It is not obvious how to choose the true energy-momentum
tensor. This problem is solved in the framework of the
multimomentum Hamiltonian formalism which provides the adequate
description of constraint field systems. The goal is that, for
different solutions of the same constraint field model, one should
choose different stress-energy-momentum tensors in general. Gauge
theory illustrates this result. The stress-energy-momentum tensors of
affine-metric gravity are examined.
\end{abstract}

\section{Introduction}

In analytical mechanics, there exists the conventional energy
conservation law, otherwise in field theory. Here, we are concerned
only with differential conservation laws.

Let $F$ be a manifold. In time-dependent mechanics on the phase
space ${\bf R}\times T^*F$ coordinatized by $(t,y^i,\dot
y_i)$ and on the configuration space ${\bf R}\times TF$ coordinatized by
$(t,y^i,\dot y^i)$, the Lagrangian energy and the construction of
the Hamiltonian formalism require the prior choice of a connection on
the bundle ${\bf R}\times F\to {\bf R}$ \cite{ech}. Such a connection
however is usually hidden by utilizing the natural trivial connection
on this bundle. Given  a
Hamiltonian function $\cH$ on the phase space ${\bf R}\times T^*F$,
we therefore have the conventional energy conservation law
\begin{equation}
\frac{d\cH}{dt}\ap \frac{\dr\cH}{\dr t} \label{E1}
\end{equation}
where by "$\ap$" is meant the weak identity modulo the Hamilton equations.
Given a Lagrangian function $\cL$ on the configuration space  ${\bf
R}\times TF$, there exists the fundamental identity
\begin{equation}
\frac{\dr\cL}{\dr t}+
\frac{d}{dt}(\dot y^i(t)\frac{\dr\cL}{\dr\dot y^i}-\cL)\ap 0 \label{J1}
\end{equation}
modulo the motion equations. It is the energy conservation
law in the following sense.

Let $\wh L$ be the Legendre morphism
\[\dot y_i\circ\wh L=\frac{\dr\cL}{\dr\dot y^i}\]
and $Q=\Im\wh L$ the Lagrangian constraint space.
Let $\cH$ be a Hamiltonian function associated with $\cL$ and $\wh H$
the momentum morphism
\[\dot y^i\circ\wh H=\frac{\dr\cH}{\dr\dot y_i}.\]
Every solution $r$ of the Hamilton equations of $\cH$ which lives on $Q$
yields the solution $\wh H\circ r$ of the
Euler-Lagrange equations of $\cL$. Then, the identity (\ref{J1}) on $\wh
H\circ r$ recovers the energy tranformation law (\ref{E1}) on $r$.

Note that there are different Hamiltonian functions associated
with the same singular Lagranguan function as a rule. Given such a
Hamiltonian function, the Lagrangian constraint space $Q$ plays the
role of the primary constraint space, and the Dirac-Bergmann procedure
can be utilized in order to get the final constraint space where a solution of
the Hamilton equations exists \cite{leo2,2got}.

In field theory, classical fields are descirbed by
sections of a fibered manifold $Y\to X.$ Their dynamics is
phrased in terms of jet manifolds.

As a shorthand, one can say that the
$k$-order jet manifold $J^kY$ of a fibered manifold $Y\to X$
comprises the equivalence classes
$j^k_xs$, $x\in X$, of sections $s$ of $Y$ identified by the first $(k+1)$
terms of their Taylor series at a point $x$.

We restrict ourselves
to the first order  Lagrangian formalism when the configuration space
is $J^1Y$. Given fibered coordinates $(x^\m,y^i)$
of $Y$, the jet manifold $J^1Y$ is endowed with the adapted  coordinates
$ (x^\m,y^i,y^i_\m)$.
A first order Lagrangian density on $J^1Y$ is defined to be the morphism
\be
&& L: J^1Y\to\op\w^nT^*X, \qquad n=\dim X,\\
&&L=\cL(x^\m,y^i,y^i_\m)\om, \qquad \om=dx^1\w ...\w dx^n.
\ee
The corresponding first order Euler-Lagrange equations for sections
$\ol s$ of $J^1Y\to X$ read
\ben &&\dr_\la\ol s^i=\ol s^i_\la, \nonumber\\
&& \dr_i\cL-(\dr_\la+\ol s^j_\la\dr_j
+\dr_\la\ol s^j_\m\dr^\m_j)\dr^\la_i\cL=0. \label{306}\een

We consider the Lie derivatives of Lagrangian densities
in order to obtain differential conservation laws.

Let
\[ u=u^\m(x)\dr_\m + u^i(y)\dr_i\]
be a projectable vector field on $Y\to X$ and $\ol u$ its jet lift
(\ref{1.21}) onto $J^1Y\to X$. Given $L$, let us computer the Lie derivative
${\bf L}_{\ol u}L$. We get the identity
\begin{equation}
\ol s^*{\bf L}_{\ol u}L\ap -\frac{d}{dx^\la}[\pi^\la_i(u^\m \ol
s^i_\m-u^i) -u^\la\cL ]\om, \qquad \pi^\m_i=\dr^\m_i\cL, \label{502}
\end{equation}
modulo the Euler-Lagrange equations (\ref{306}).

In particular, if $u$ is a vertical vector field
this identity comes to the current conservation law exemplified by the
Noether identities in gauge theory \cite{3sar'}.

Let now $\tau=\tau^\la\dr_\la$ be a vector field on $X$ and
\begin{equation}
\tau_\G=\tau^\m (\dr_\m+\G^i_\m\dr_i)\label{J8}
\end{equation}
its horizontal lift onto $Y$ by a connection $\G$
on $Y\to X$. In this case, the identity (\ref{502}) takes the form
\begin{equation}
\ol s^*{\bf L}_{\ol\tau_\G}L\ap
-\frac{d}{dx^\la}[\tau^\m \cT_\G{}^\la{}_\m (\ol s)]\om \label{504}
\end{equation}
where
\[\cT_\G{}^\la{}_\m (\ol s) =[\pi^\la_i(y^i_\m -\G^i_\m)
-\dl^\la_\m\cL]\circ \ol s^i_\m\]
is the stress-energy-momentum (SEM)
tensor on a field $\ol s$ relative to the connection $\G$.
We here restrict ourselves to this particular case of
SEM tensors \cite{fer,got2,kij}.

For instance, let us choose the trivial local connection $\G^i_\m=0$.
In this case, the identity (\ref{504}) recovers the well-known conservation law
\[\frac{\dr\cL}{\dr x^\la} +\frac{d}{dx^\la} \cT^\la{}_\m (\ol s)\ap 0\]
of the canonical energy-momentum tensor
\begin{equation}
\cT^\la{}_\m (\ol s)= \pi^\la_i\ol s^i_\m -\dl^\la_\m\cL. \label{J6}
\end{equation}
Physicists lose sight of the fact that (\ref{J6}) fails to be a mathematical
well-behaved object.

The crucial point lies in the fact that the Lie derivative
\[{\bf L}_{\ol\tau_\G}L=\{\dr_\m\tau^\m\cL +[\tau^\m\dr_\m
+\tau^\m\G^i_\m\dr_i+(\dr_\la(\tau^\m\G^i_\m)+\tau^\m
y^j_\la\dr_j\G^i_\m -y^i_\m\dr_\la\tau^\m)\dr^\la_i]\cL\}\om \]
is almost never equal to zero. Therefore, it
is not obvious how to choose the true energy-momentum tensor.

For instance, the canonical energy-momentum tensor (\ref{J6}) in gauge
theory is symmetrized by hand in order to obtain the gauge invariant
one. In gauge theory in the presence of a background
world metric $g$, the identity
(\ref{504}) is brought into the covariant conservation law
\begin{equation}
\nabla_\la t^\la{}_\m\ap 0 \label{E2}
\end{equation}
for the metric energy-momentum tensor.

In Einstein's General Relativity, the covariant conservation law (\ref{E2})
issues directly from the gravitation equations.
It is concerned with the zero-spin matter in the
presence of the gravitational field generated by this matter,
though the matter is not required to satisfy the motion equations.
The total energy-momentum conservation law for matter and gravity is
introduced by hand. It is written
\begin{equation}
\frac{d}{dx^\m}[(-g)^N(t^{\la\m} +t_g{}^{\la\m)})]\ap 0\label{E3}
\end{equation}
where the energy-momentum pseudotensor $t_g{}^{\la\m}$ of a metric
gravitational field is defined to satisfy the relation
\[(-g)^N(t^{\la\m} +t_g{}^{\la\m})\ap \frac{1}{2\kp}\dr_\si\dr_\al
[(-g)^N(g^{\la\m} g^{\si\al}-g^{\si\m} g^{\la\al}]\]
modulo the Einstein equations. However, the conservation law (\ref{E3})
appears satisfactory only in cases of asymptotic-flat gravitational
fields and of a background metric.

Moreover, the covariant conservation law (\ref{E2}) fails to take place in the
affine-metric gravitation theory and in the gauge gravitation theory,
e.g., in the presence of fermion fields.

Thus, we have not any conventional energy-momentum
conservation law in Lagrangian field theory. In particular, one may
take different SEM tensors for different field models and, moreover,
different SEM tensors for different solutions of the same field equations.
Just the latter in fact is the above-mentioned symmetrization of the
canonical energy-momentum tensor in gauge theory.

Gauge theory exemplifies constraint field theories. Contemporary field
models are almost always the constraint ones. To describe them, let us
turn to the Hamiltonian formalism.

When applied to field theory, the conventional Hamiltonian formalism
takes the form of the instantaneous
Hamiltonian formalism where canonical variables are field functions at a
given instant of time. The corresponding phase space is
infinite-dimensional, so that the Hamilton equations in the bracket form fail
to be differential equations.

The true partners of the Lagrangian formalism in classical field theory
are polysymplectic and multisymplectic Hamiltonian machineries where
canonical momenta correspond to derivatives of fields with respect to all world
coordinates, not only the temporal one \cite{car,gun,6sar,sard}.
We here follow the multimomentum Hamiltonian formulation
of field theory when the phase space of fields is the Legendre bundle
\begin{equation}
\Pi=\op\w^n T^*X\op\ot_Y TX\op\ot_Y V^*Y \label{00}
\end{equation}
over $Y$ which is coordinatized by
$(x^\la ,y^i,p^\la_i)$  \cite{sard,7sar,Esar}. Every
Lagrangian density  $L$ on $J^1Y$ implies the Legendre morphism
\be
&&\wh L:J^1Y\op\to_Y \Pi,\\
&& p^\m_i\circ\wh L=\pi^\m_i.
\ee

The Legendre bundle (\ref{00}) carries the polysymplectic form
\begin{equation}
\Om =dp^\la_i\w
dy^i\w\om\ot\dr_\la. \label{406}
\end{equation}
We say that a connection $\g$ on the fibered Legendre manifold $\Pi\to
X$ is a Hamiltonian connection if the form $\g\rfloor\Om$ is closed.
Then, a Hamiltonian form $H$ on $\Pi$ is defined to be an
exterior form such that
\begin{equation}
dH=\g\rfloor\Om \label{013}
\end{equation}
for some Hamiltonian connection $\g$. The key point lies in the fact
that every Hamiltonian form admits splitting
\begin{equation}
H =p^\la_idy^i\w\om_\la -p^\la_i\G^i_\la\om
-\wt{\cH}_\G\om=p^\la_idy^i\w\om_\la-\cH\om,\qquad
\om_\la=\dr_\la\rfloor\om,  \label{017}
\end{equation}
where $\G$ is a connection on $Y\to X$.

Given the splitting (\ref{017}), the equality
(\ref{013}) comes to the Hamilton equations
\ben
&&\dr_\la r^i =\dr^i_\la\cH, \nonumber\\
&&\dr_\la r^\la_i=-\dr_i\cH \label{3.11}
\een
for sections $r$ of $\Pi\to X$.

The Hamilton equations (\ref{3.11}) are the multimomentum generalization
of the Hamilton equations in mechanics. The corresponding
multimomentum generalization of the conventional energy conservation law
(\ref{E1}) is the weak identity
\begin{equation}
\tau^\m[(\dr_\m+\G^i_\m\dr_i-\dr_i\G^j_\m
r^\la_j\dr^i_ \la)\wt{\cH}_\G-\frac{d}{dx^\la}
T_\G{}^\la{}_\m (r)]\ap\tau^\m r^\la_iR^i_{\la\m}, \label{5.27}
\end{equation}
\begin{equation}
T_\G{}^\la{}_\m (r)=[r^\la_i\dr^i_\m\wt{\cH}_\G-\dl^\la_\m(r^\al_i\dr^i_\al
\wt{\cH}_\G -\wt{\cH}_\G)] \label{J3}
\end{equation}
where
\[ R^i_{\la\m} =\dr_\la\G^i_\m -\dr_\m\G^i_\la +\G^j_\la\dr_j\G^i_\m
-\G^j_\m\dr_j\G^i_\la\]
is the curvature of the connection $\G$. One can think of the tensor
(\ref{J3}) as being the Hamiltonian SEM tensor.

If a Lagrangian density is regular, the
multimomentum Hamiltonian formalism is equivalent to
the Lagrangian formalism, otherwise in case of degenerate
Lagrangian densities.
In field theory, if a Lagrangian density is not regular,
the Euler-Lagrange equations become
underdetermined and require supplementary gauge-type conditions.
In gauge theory, they are the familiar gauge conditions.
In general case, the
gauge-type conditions however remain elusive. In the framework of
the multimomentum Hamiltonian formalism, they appear automatically
as a part of the Hamilton equations.
The key point consists in the fact that, given a degenerate
Lagrangian density, one must consider a family of different associated
Hamiltonian forms in order to
exaust all solutions of the Euler-Lagrange equations.

Lagrangian densities of all realistic field models are
quadratic or affine in the velocity coordinates $y^i_\m$. Complete
family of Hamiltonian forms associated with such a Lagrangian density
always exists \cite{sard,7sar,Esar}. Moreover, these Hamiltonian forms
differ from each other only in connections $\G$ in the splitting
(\ref{017}). Different connections are responsible for different gauge-type
conditions mentioned above. They are also the connections which one
should utilize in construction of the Hamiltonian SEM tensors (\ref{J3}).

Thus, the tools are now at hand in order to examine
the energy-momentum conservation laws in constraint field models.

As a test case, we construct Hamiltonian SEM tensors in gauge theory
and then in gravitation theory.

The identity  (\ref{5.27}) remains true in
the first order Lagrangian theories of gravity.
In this work, we examine the metric-affine gravity where
independent dynamic variables are world metrics and general linear
connections. The energy-momentum conservation law
in the affine-metric gravitation theory is not widely discussed
\cite{heh,E2sar}. We construct the Hamiltonian SEM tensor for gravity.
In case of the affine Hilbert-Einstein Lagrangian density, it is equal to
\[T^\la_\m=\frac{1}{2\kp}\dl^\la_\m R\sqrt{-g}\]
and the total conservation law (\ref{5.27}) for matter
and gravity is reduced to the conservation law for matter in the
presence of a background world metric, otherwise in case of quadratic
Lagrangian densities.

\section{Technical preliminary}

The first order jet manifold $J^1Y$ of
$Y$ is both the fibered manifold $J^1Y\to X$
and the affine bundle $J^1Y\to Y $  modelled on the vector
bundle $T^*X\ot_Y VY.$ Hereafter, $J^1Y$ is identified to its
image under the contact map
\ben &&\la:J^1Y\op\to_YT^*X \op\ot_Y TY,\nonumber\\
&&\la=dx^\la\ot(\dr_\la+y^i_\la \dr_i).\label{18}\een

Recall that every fibered morphism of $\Phi: Y\to Y'$
over a diffeomorphism of $X$ has the jet prolongation to the fibered morphism
\[J^1\Phi:J^1Y\to J^1Y',\]
\[ {y'}^i_\m\circ
J^1\Phi=(\dr_\la\Phi^i+\dr_j\Phi^iy^j_\la)\frac{\dr x^\la}{\dr {x'}^\m}.\]

Every projectable vector field
\[u = u^\la\dr_\la + u^i\dr_i\]
on $Y\to X$ gives rise to the projectable vector field
\begin{equation}
\ol u =u^\la\dr_\la + u^i\dr_i + (\dr_\la u^i+y^j_\la\dr_ju^i
- y_\m^i\dr_\la u^\m)\dr_i^\la, \label{1.21}
\end{equation}
on $J^1Y\to X$.

The contact map (\ref{18}) implies the bundle monomorphism
\[ \wh\la: J^1Y\op\times_X TX\ni\dr_\la\mapsto \dr_\la\rfloor\la
\in J^1Y\op\times_Y TY\]
and the canonical horizontal splitting of the pullback
\[J^1Y\op\times_Y TY=\wh\la(TX)\op\oplus_{J^1Y} VY,\]
\[\dot x^\la\dr_\la
+\dot y^i\dr_i =\dot x^\la(\dr_\la +y^i_\la\dr_i) + (\dot y^i-\dot x^\la
y^i_\la)\dr_i.\]
Building on this splitting,
one obtains the corresponding horizontal splittings
\begin{equation}
u =u^\la\dr_\la +u^i\dr_i=u_H +u_V =u^\la (\dr_\la +y^i_\la
\dr_i)+(u^i - u^\la y^i_\la)\dr_i. \label{31}
\end{equation}
of the pullback of a projectable vector field on $Y\to X$ onto $J^1Y$.

There is the 1:1 correspondence between the global sections
\[\G =dx^\la\ot(\dr_\la+\G^i_\la\dr_i)\]
of the affine jet bundle $J^1Y\to Y$ and the connections
on $Y\to X$. These connections constitute the
affine space modelled on the linear space of soldering forms
$Y\to T^*X\op\ot_YVY$ on $Y$.

The repeated jet manifold
$J^1J^1Y$, by definition, is the first order jet manifold of
$J^1Y\to X$. It is provided with the adapted coordinates
$(x^\la ,y^i,y^i_\la ,y_{(\m)}^i,y^i_{\la\m})$.
Its subbundle $ \wh J^2Y$ given by the coordinate relation
$y^i_{(\la)}= y^i_\la$ is called the sesquiholonomic jet manifold.
The second order jet manifold $J^2Y$ of $Y$ is the subbundle
of $\wh J^2Y$ where $ y^i_{\la\m}= y^i_{\m\la}.$

\section{SEM tensors in the Lagrangian formalism}

Given a Lagrangian density $L$, the jet manifold $J^1Y$  carries
the associated Poincar\'e-Cartan form
\begin{equation}
\Xi_L=\pi^\la_idy^i\w\om_\la -\pi^\la_iy^i_\la\om +\cL\om \label{303}
\end{equation}
and the Lagrangian  polysymplectic form
\[\Om_L=(\dr_j\pi^\la_idy^j+\dr^\m_j\pi^\la_idy^j_\m)\w
dy^i\w\om\ot\dr_\la.\]
Using the pullback of these forms onto the
repeated jet manifold $J^1J^1Y$, one can construct the exterior generating form
\begin{equation}
\La_L=d\Xi_L-\la\rfloor\Om_L=[y^i_{(\la)}-y^i_\la)d\pi^\la_i +
(\dr_i-\wh\dr_\la\dr^\la_i)\cL dy^i]\w\om,\label{304}
\end{equation}
\[ \la=dx^\la\ot\wh\dr_\la,\qquad
\wh\dr_\la =\dr_\la +y^i_{(\la)}\dr_i+y^i_{\m\la}\dr^\m_i,\]
on $J^1J^1Y$.
Its restriction to the sesquiholonomic jet manifold
$\wh J^2Y$ defines the first order Euler-Lagrange operator
\ben
&&\cE'_L:\wh J^2Y\op\to_{J^1Y}\op\w^{n+1}T^*Y,\nonumber\\
&&\cE_L=\delta_i\cL dy^i\w\om=
 [\dr_i-(\dr_\la +y^i_\la\dr_i+y^i_{\m\la}\dr^\m_i)
\dr^\la_i]\cL dy^i\w\om ,\label{305}
\een
corresponding to $L$. The restriction of the form (\ref{304})
to the second order jet manifold $J^2Y$ of $Y$ recovers
the familiar variational Euler-Lagrange operator
\[\cE_L: J^2Y\op\to_Y\op\w^{n+1}T^*Y.\]
It is given by the expression
(\ref{305}), but with symmetric coordinates $y^i_{\m\la}=y^i_{\la\m}$.

Let $\ol s$ be a section of the fibered jet manifold $J^1Y\to X$ such that
its jet prolongation  $J^1\ol s$ takes its values into
$\Ker\cE'_L$ given by the coordinate relations
\[ \dr_i\cL-(\dr_\la+y^j_\la\dr_j +y^j_{\m\la}\dr^\m_j)\dr^\la_i\cL=0.\]
Then, $\ol s$ satisfies the first order
Euler-Lagrange equations (\ref{306}).
These equations  are equivalent to the second order Euler-Lagrange equations
\begin{equation}
\dr_i\cL-(\dr_\la+\dr_\la s^j\dr_j
+\dr_\la\dr_\m s^j \dr^\m_j)\dr^\la_i\cL=0.\label{2.29}
\end{equation}
for sections $s$ of $Y\to X$ where $\ol s=J^1s$.

We have the following differential conservation laws on
solutions of the first order Euler-Lagrange equations.

Given a Lagrangian density $L$ on $J^1Y$, let us consider its pullback onto
$\wh J^2Y$. Let $u$
be a projectable vector field on $Y\to X$ and $\ol u$ its jet lift
(\ref{1.21}) onto $J^1Y\to X$. Its pullback onto
$J^1J^1Y$ has the
the canonical horizontal splitting (\ref{31}) given by the expression
\[\ol u=\ol u_H+\ol u_V = u^\la(\dr_\la+ y^i_{(\la)}\dr_i+y^i_{\m\la}\dr^\m_i)
+[(u^i-y^i_{(\la)} u^\la)\dr_i+(u^i_\m- y^i_{\m\la}u^\la)\dr^\m_i].\]

Let us compute the Lie derivative ${\bf L}_{\ol u}L$. We have
\begin{equation}
{\bf L}_{\ol u}L= \wh \dr_\la[\pi^\la_i(u^i-u^\m y^i_\m ) +u^\la\cL]\om
+\ol u_V\rfloor\cE'_L, \label{501}
\end{equation}
\[\wh\dr_\la =\dr_\la +y^i_\la\dr_i+y^i_{\m\la}\dr^\m_i.\]
Being restricted to $\Ker\cE'_L$, the equality (\ref{501}) is written
\begin{equation}
\dr_\la u^\la\cL +[u^\la\dr_\la+
u^i\dr_i +(\dr_\la u^i +y^j_\la\dr_ju^i -y^i_\m\dr_\la u^\m)\dr^\la_i]\cL
\ap \wh \dr_\la[\pi^\la_i(u^i-u^\m y^i_\m )+u^\la\cL].\label{J4}
\end{equation}
On solutions $\ol s$ of the first order Euler-Lagrange equations, the
weak identity (\ref{J4}) comes to the differential conservation law
\[\ol s^*{\bf L}_{\ol u}L\ap d(\ol u\rfloor\Xi_L\circ\ol s) \]
which takes the coordinate form (\ref{502}).

In particular, let $\tau_\G$
be the horizontal lift (\ref{J8}) of a vector field $\tau$
on $X$ onto $Y\to X$ by a connection $\G$
on $Y$. In this case, the identity (\ref{J4}) is written
\begin{equation}
\tau^\m\{[\dr_\m
+\G^i_\m\dr_i +(\dr_\la\G^i_\m +y^j_\la\dr_j\G^i_\m)\dr^\la_i]\cL+
\wh\dr_\la [\pi^\la_i(y^i_\m-\G^i_\m)-\dl^\la_\m\cL]\}\ap 0. \label{J5}
\end{equation}
On solutions $\ol s$ of the first order Euler-Lagrange equations, the
identity (\ref{J5}) comes to the differential conservation law (\ref{504})
where $\cT_\G{}^\la{}_\m (\ol s)$ are coefficients of the $T^*X$-valued form
\begin{equation}
\cT_\G(\ol s)=-(\G\rfloor\Xi_L)\circ\ol s =[\pi^\la_i(\ol s^i_\m-\G^i_\m)
-\dl^\la_\m\cL]dx^\m\ot\om_\la \label{S14}
\end{equation}
on $X$. This conservation law takes the coordinate form
\[\tau^\m\{[\dr_\m
+\G^i_\m\dr_i +(\dr_\la\G^i_\m +\ol s^j_\la\dr_j\G^i_\m)\dr^\la_i]\cL+
\frac{d}{dx^\la} [\pi^\la_i(\ol s^i_\m-\G^i_\m)-\dl^\la_\m\cL]\}\ap 0. \]

\section{Multimomentum Hamiltonian formalism}

Let $\Pi$ be the Legendre bundle (\ref{00})
coordinatized by $( x^\la ,y^i,p^\la_i)$.
By $J^1\Pi$ is meant the first order jet manifold of
$\Pi\to X$. It is coordinatized by
$( x^\la ,y^i,p^\la_i,y^i_{(\m)},p^\la_{i\m})$.

The Legendre manifold $\Pi$ carries the generalized Liouville form
\[\th =-p^\la_idy^i\w\om\ot\dr_\la \]
and the polysymplectic form $\Om$ (\ref{406}).

The Hamiltonian formalism in fibered manifolds is formulated
intrinsically in terms of Hamiltonian connections which play the
role similar to that of Hamiltonian vector fields in the symplectic
geometry \cite{sard,7sar,Esar}.

A connection
\[\g=dx^\m\ot(\dr_\m +\g^i_{(\m)}\dr_i +\g^\la_{i\m}\dr^i_\la)\]
on the fibered Legendre manifold $\Pi\to X$, by definition above,
is called the Hamiltonian
connection if the exterior form $\g\rfloor\Om$  is closed.
An exterior $n$-form $H$ on $\Pi$ is called a  Hamiltonian form if
there exists a Hamiltonian connection  satisfying the equation (\ref{013}).
Hamiltonian forms always exist as follows.

Every connection $\G$ on $Y\to X$ gives rise to the connection
\[\wt\G =dx^\la\ot[\dr_\la +\G^i_\la\dr_i +
(-\dr_j\G^i_\la p^\m_i-K^\m{}_{\n\la}p^\n_j+K^\al{}_{\al\la}
p^\m_j)\dr^j_\m] \]
on $\Pi\to X$ where
\[K^*=dx^\la\ot(\dr_\la +K^\m{}_{\n\la}\dot x_\m\frac{\dr}{\dr\dot x_\n})\]
is a linear symmetric connection on $T^*X$. We have the equality
\[\wt\G\rfloor\Om =d(\G\rfloor\th).\]
A glance at this equality shows that $\wt\G$ is a Hamiltonian connection and
\[ H_\G=\G\rfloor\th =p^\la_i dy^i\w\om_\la -p^\la_i\G^i_\la\om\]
is a Hamiltonian form.

\begin{lemma}
Let $H$ be a Hamiltonian form. For any exterior horizontal density
$\wt H=\wt{\cH}\om$ on $\Pi\to X$, the form $H+\wt H$ is a Hamiltonian form.
Conversely, if $H$ and $H'$ are  Hamiltonian forms,
their difference $H-H'$ is an exterior horizontal density on $\Pi\to X$.
\end{lemma}

Thus, Hamiltonian  forms constitute an affine space
modelled on a linear space of the exterior horizontal densities on
$\Pi\to X$. It follows that every Hamiltonian form on $\Pi$ can be
given by the expression (\ref{017}) where $\G$ is some connection on $Y\to X$.
Moreover, a Hamiltonian form has the canonical splitting (\ref{017})
as follows.

Every  Hamiltonian form $H$ implies the momentum morphism
\be
&& \wh H:\Pi\op\to_Y J^1Y, \\
&& y_\la^i\circ\wh H=\dr^i_\la\cH,
\ee
and the associated connection $\G_H =\wh H\circ\wh 0$
on $Y$ where $\wh 0$ is the global zero section of $\Pi\to Y$.
As a consequence, we have the canonical splitting \[ H=H_{\G_H}-\wt H.\]

The Hamilton operator $\cE_H$ of a Hamiltonian form $H$
is defined to be the first order differential operator
\ben
&& \cE_H :J^1\Pi\to\op\w^{n+1} T^*\Pi,\nonumber \\
&& \cE_H=dH-\wh\Om=[(y^i_{(\la)}-\dr^i_\la\cH) dp^\la_i
-(p^\la_{i\la}+\dr_i\cH) dy^i]\w\om \label{3.9}
\een
on $\Pi\to X$ where
\[\wh\Om=dp^\la_i\w
dy^i\w\om_\la +p^\la_{i\la}dy^i\w\om -y^i_{(\la)} dp^\la_i\w\om\]
is the pullback of the multisymplectic form (\ref{406}) onto $J^1\Pi$.

For any connection $\g$ on $\Pi\to X$, we have
\[\cE_H\circ\g =dH-\g\rfloor\Om.\]
It follows that  $\g$  is a Hamiltonian connection for a
Hamiltonian form $H$ iff it takes its values into
$\Ker\cE_H$ given by the coordinate relations
\begin{equation}
y^i_{(\la)} =\dr^i_\la\cH, \qquad p^\la_{i\la}=-\dr_i\cH. \label{3.10}
\end{equation}

Let a Hamiltonian connection $\g$ has an integral section $r$ of
$\Pi\to X$, that is, $\g\circ r=J^1r$.
Then, the algebraic equations (\ref{3.10}) are brought into the first
order differential Hamilton equations (\ref{3.11}).

Now we consider relations between Lagrangian and Hamiltonian formalisms.
A  Hamiltonian form
$H$ is defined to be associated with a Lagrangian density $L$ if it
satisfies the relations
\bea &&\wh L\circ\wh H|_Q=\Id_Q, \qquad Q=\wh L( J^1Y) \label{2.30a},\\
&& H=H_{\wh H}+L\circ\wh H \label{2.30b}\eea
which take the coordinate form
\[\dr^\m_i\cL(x^\la, y^j, \dr^j_\la\cH)= p^\m_i,\]
\[\cL(x^\la, y^j, \dr^j_\la\cH)=p^\m_i\dr^i_\m\cH -\cH.\]
Note that there are different Hamiltonian forms associated with the same
singular Lagrangian density as a rule.

Bearing in mind physical application, we restrict our consideration to
so-called semiregular Lagrangian densities
$L$ when the preimage $\wh L^{-1}(q)$ of each point of
$q\in Q$ is the connected submanifold of $J^1Y$.
In this case, all Hamiltonian forms associated
with a semiregular Lagrangian density $L$ coincide on the
Lagrangian constraint space $Q$,
and the Poincar\'e-Cartan form $\Xi_L$ is the pullback
\be
&&\Xi_L=H\circ\wh L,\\
&&\pi^\la_iy^i_\la-\cL=\cH(x^\m,y^i,\pi^\la_i),
\ee
of any associated multimomentum Hamiltonian form $H$ by the Legendre
morphism $\wh L$ \cite{zak}. Also the generating form
(\ref{304}) is the pullback of
\[\La_L=\cE_H\circ J^1\wh L\]
of the Hamilton operator (\ref{3.9}) of any Hamiltonian form $H$
associated with a semiregular Lagrangian density $L$. As a consequence,
we obtain the following correspondence between solutions of the Euler-Lagrange
equations and the Hamilton equations \cite{7sar,zak}.

Let a  section $r$ of $\Pi\to X$
be a solution of the Hamilton equations (\ref{3.11})
for a Hamiltonian form $H$ associated with a semiregular Lagrangian
density $L$. If $r$ lives on the Lagrangian constraint space $Q$, the section
$\ol s=\wh H\circ r$ of $J^1Y\to X$ satisfies the first
order Euler-Lagrange equations (\ref{306}).
Conversely, given a semiregular Lagrangian density $L$, let
$\ol s$ be a solution of the
first order Euler-Lagrange equations (\ref{306}).
Let $H$ be a Hamiltonian form associated with $L$ so that
\begin{equation}
\wh H\circ \wh L\circ \ol s=\ol s.\label{2.36}
\end{equation}
Then, the section $r=\wh L\circ \ol s$ of $\Pi\to X$ is a solution of the
Hamilton equations (\ref{3.11}) for $H$.
For sections $\ol s$ and $r$, we have the relations
\[\ol s=J^1s, \qquad  s=\pi_{\Pi Y}\circ r\]
where $s$ is a solution of the second order Euler-Lagrange equations
(\ref{2.29}).

We shall say that a family of Hamiltonian forms $H$
associated with a semiregular Lagrangian density $L$ is
complete if, for each solution $\ol s$ of the first order Euler-Lagrange
equations (\ref{306}), there exists
a solution $r$ of the Hamilton equations (\ref{3.11}) for
some  Hamiltonian form $H$ from this family so that
\begin{equation}
r=\wh L\circ\ol s,\qquad  \ol s =\wh H\circ r, \qquad
\ol s= J^1(\pi_{\Pi Y}\circ r). \label{2.37}
\end{equation}
Such a complete family
exists iff, for each solution $\ol s$ of the Euler-Lagrange
equations for $L$, there exists a  Hamiltonian form $H$ from this
family so that the condition (\ref{2.36}) holds.

We do not discuss here existence of solutions of Euler-Lagrange and
Hamilton equations \cite{leo1}. Note that, in cotrast with mechanics,
there are different Hamiltonian connections associated with the same
multimomentum Hamiltonian form in general. Moreover, in field theory
when the primary constraint space is the Lagrangian constraint space $Q$,
there is a family of Hamiltonian forms associated with the same
Lagrangian density as a rule. In practice, one can choose either
the Hamilton equations or solutions of the Hamilton equations such
that these solutions live on the constraint space.

\section{Hamiltonian SEM tensors}

Let $H$ be a Hamiltonian form on the Legendre bundle $\Pi$
over a fibered manifold $Y\to X$.
We have the following differential conservation law on solutions of the
Hamilton equations.

Let $r$ be a section of the fibered Legendre manifold $\Pi\to X$.
Given a connection $\G$ on $Y\to X$, we consider the
$T^*X$-valued $(n-1)$-form
\ben &&T_\G(r)=-(\G\rfloor H)\circ r,\nonumber\\
&&T_\G (r)=[r^\la_i(\dr_\m r^i -\G^i_\m)-\dl^\la_\m(r^\al_i(\dr_\al r^i
-\G^i_\al) -\wt{\cH}_\G)] dx^\m\ot\om_\la, \label{5.8}\een
on $X$ where $\wt{\cH}_\G$ is the Hamiltonian density in the splitting
(\ref{017}) of $H$ with respect to the connection $\G$.

Let $\tau=\tau^\la\dr_\la$
be a vector field on $X$. Given a connection $\G$ on $Y\to X$, it
gives rise to the projectable vector field
\[\wt\tau_\G= \tau^\la\dr_\la + \tau^\la\G^i_\la\dr_i +
(-\tau^\m p^\la_j\dr_i\G^j_\m
-p^\la_i\dr_\m\tau^\m + p^\m_i\dr_\m\tau^\la) \dr^i_\la\]
on the Legendre bundle $\Pi$. Let us calculate the Lie derivative
${\bf L}_{\wt\tau_\G}\wt H_\G$ on a section $r$. We have
\ben
&&({\bf L}_{\wt\tau_\G}\wt H_\G)\circ r=
\{\dr_\la\tau^\la\wt{\cH}_\G +[\tau^\la\dr_\la + \tau^\la\G^i_\la\dr_i +
(-\tau^\m r^\la_j\dr_i\G^j_\m
-r^\la_i\dr_\m\tau^\m + r^\m_i\dr_\m\tau^\la) \dr^i_\la]\wt{\cH}_\G\}\om
\nonumber\\
&&\qquad =\tau^\m r^\la_iR^i_{\la\m}\om +d(\tau^\m
T_\G{}^\la{}_\m (r)\om_\la)-(\wt\tau_{\G V}\rfloor\cE_H)\circ r\label{221}
\een
where $\wt\tau_{\G V}$ is the vertical part of the canonical
horizontal splitting (\ref{31}) of the vector field $\wt\tau_V$ on $\Pi$
over $J^1\Pi$. If $r$ is a
solution of the Hamilton equations, the equality (\ref{221}) comes
to the conservation law (\ref{5.27}).
The form (\ref{5.8}) modulo the Hamilton equations reads
\begin{equation}
T_\G(r)\ap [r^\la_i(\dr^i_\m\cH -\G^i_\m)-
\dl^\la_\m(r^\al_i\dr^i_\al\cH-\cH)]dx^\m\ot\om_\la.\label{5.26}
\end{equation}

For instance, if $X={\bf R}$ and $\G$ is the trivial connection, we have
\[T_\G(r)=\cH dt\]
where $\cH$ is a Hamiltonian function. Then, the identity
(\ref{5.27}) comes to the conventional energy conservation law (\ref{E1})
in mechanics.

Unless $n=1$, the identity (\ref{5.27}) can not be regarded directly
as the energy-momentum conservation law. To clarify its physical meaning,
we turn to the Lagrangian formalism.

\begin{lemma} Let a Hamiltonian form $H$ be associated with a
semiregular Lagrangian density $L$. Let $r$ be a solution
of the Hamilton equations of $H$ which lives on the Lagrangian
constraint space $Q$ and $\ol s$ the associated solution  of the first order
Euler-Lagrange equations of $L$ so that they satisfy the conditions
(\ref{2.37}). Then, we have
\be
&&T_\G (r)=\cT_\G(\wt H\circ r),\\
&&T_\G (\wt L\circ \ol s) =\cT_\G (\ol s)
\ee
where $\cT_\G$ is the SEM tensor (\ref{S14}).
\end{lemma}

It follows that, on the Lagrangian constraint space $Q$, the form
(\ref{5.26}) can be treated the Hamiltonian
SEM tensor relative to the connection $\G$ on $Y\to X$.

At the same time, the examples below show
that, in several field models, the equality (\ref{5.27}) is brought into the
covariant conservation law (\ref{E2}) for the metric energy-momentum tensor.

In the Lagrangian formalism, the metric
energy-momentum tensor is defined to be
\[ \sqrt{-g} t_{\al\bt}=2\frac{\dr\cL}{\dr g^{\al\bt}}.\]
In case of a background world metric $g$, this object is well-behaved.
In the framework of the multimomentum Hamiltonian formalism,
one can introduce the similar tensor
\begin{equation}
\sqrt{-g}t_H{}^{\al\bt}=2\frac{\dr\cH}{\dr g_{\al\bt}}.\label{5.28}
\end{equation}

If a Hamiltonian form $H$ is associated
with a semiregular Lagrangian density $L$, there are the equalities
\be
&&t_H{}^{\al\bt}(q)
=-g^{\al\m}g^{\bt\n}t_{\m\n}(x^\la,y^i,\dr_\la^i\cH(q)), \qquad q\in Q,\\
&&  t_H{}^{\al\bt}(x^\la,y^i,\pi^\la_i(z))=-g^{\al\m}g^{\bt\n}t_{\m\n}(z),
\qquad \wh H\circ\wh L(z)=z.
\ee
In view of these equalities, we can think of the tensor (\ref{5.28})
restricted to the Lagrangian constraint space $Q$ as being the
Hamiltonian metric energy-momentum tensor.
On $Q$, the tensor (\ref{5.28}) does not depend upon choice of
a Hamiltonian form $H$ associated with $L$. Therefore, we shall
denote it by the common symbol $t$. Set
\[t^\la{}_\al = g_{\al\n}t^{\la\n}.\]
In the presence of a background world metric
$g$, the identity (\ref{5.27}) takes the form
\begin{equation}
t^\la{}_\al\{^\al{}_{\la\m}\}\sqrt{-g}
+(\G^i_\m\dr_i-\dr_i\G^j_\m r^\la_j\dr^i_
\la)\wt{\cH}_\G\ap\frac{d}{dx^\la} T_\G{}^\la{}_\m +r^\la_iR^i_{\la\m}
\label{5.29}
\end{equation}
where by $\{^\al{}_{\la\m}\}$ are meant the Cristoffel symbols of the
world metric $g$.

\section{SEM tensors in gauge theory}

As a test case, let us consider the gauge
theory of principal connections treated the gauge potentials.

In the rest of the article, the manifold $X$ is assumed to be
oriented and provided with a nondegenerate fibre metric $g_{\m\n}$
in the tangent bundle of $X$. We denote $g=\det(g_{\m\n}).$

Let $P\to X$ be a principal bundle with a structure
Lie group $G$ which acts freely and transitively on $P$ on the right:
\[r_g : p\mapsto pg, \qquad  p\in P,\quad g\in G.\label{1}\]
A principal connection $A$ on
$P\to X$ is defined to be a $G$-equivariant connection on $P$ such that
\[J^1r_g\circ A= A\circ r_g\]
for each canonical morphism (\ref{1}).
There is the 1:1 correspondence between the principal connections on a
principal bundle $P\to X$  and the global sections of the quotient bundle
\begin{equation}
C:=J^1 P/G\to X. \label{68}
\end{equation}
The bundle (\ref{68}) is the affine bundle modelled on the vector bundle
\[\ol C =T^*X \ot (VP/G).\]

Given a bundle atlas $\Psi^P$ of $P$, the bundle $C$
is provided with  the fibered coordinates $(x^\m,k^m_\m)$ so that
\[(k^m_\m\circ A)(x)=A^m_\m(x)\]
are coefficients of the local connection 1-form of a principal connection
$A$ with respect to the atlas $\Psi^P$.
The first order jet manifold $J^1C$ of the bundle $C$ is
coordinatized by $(x^\m, k^m_\m, k^m_{\m\la}).$

There exists the canonical splitting
\begin{equation}
J^1C=C_+\op\oplus_C C_-=(J^2P/G)\op\oplus_C
(\op\w^2 T^*X\op\ot_C V^GP), \label{N31}
\end{equation}
\[ k^m_{\m\la}=\frac12(k^m_{\m\la}+k^m_{\la\m}+c^m_{nl}k^n_\la k^l_\m)
+\frac12( k^m_{\m\la}-k^m_{\la\m} -c^m_{nl}k^n_\la k^l_\m),\]
over $C$. The corresponding surjections are written
\be &&{\cal S}: J^1 C\to C_+, \qquad {\cal S}^m_{\la\m}=
k^m_{\m\la}+k^m_{\la\m} +c^m_{nl}k^n_\la k^l_\m,\\
&& \cF: J^1 C\to C_-,\qquad
\cF^m_{\la\m}= k^m_{\m\la}-k^m_{\la\m} -c^m_{nl}k^n_\la k^l_\m.\ee

The Legendre bundle over the bundle $C$ is
\[\Pi=\op\w^n T^*X\ot TX\op\ot_C [C\times\ol C]^*.\]
It is coordinatized by $(x^\m,k^m_\m,p^{\m\la}_m)$.

On the configuration space (\ref{N31}),
the conventional Yang-Mills Lagrangian density $L_{YM}$ of gauge
potentials in the presence of a background world metric
is given by the expression
\begin{equation}
L_{YM}=\frac{1}{4\ve^2}a^G_{mn}g^{\la\m}g^{\bt\n}\cF^m_{\la
\beta}\cF^n_{\m\n}\sqrt{| g|}\,\om \label{5.1}
\end{equation}
where  $a^G$ is a nondegenerate $G$-invariant metric
in the Lie algebra of $G$. The Legendre morphism
associated with the Lagrangian density (\ref{5.1}) takes the form
\bea &&p^{(\m\la)}_m\circ\wh L_{YM}=0, \label{5.2a}\\
&&p^{[\m\la]}_m\circ\wh L_{YM}=\ve^{-2}a^G_{mn}g^{\la\al}g^{\m\bt}
\cF^n_{\al\bt}\sqrt{| g|}. \label{5.2b}\eea
The equation (\ref{5.2a}) defines the constraint space of gauge theory.

Following the general procedure \cite{sard,7sar,Esar}, let us
consider connections on the bundle $C$ which
take their values into $\Ker\wh L_{YM}$:
\begin{equation}
\G:C\to C_+, \qquad
\G^m_{\m\la}-\G^m_{\la\m}-c^m_{nl}k^n_\la k^l_\m=0. \label{69}
\end{equation}
Moreover, we can restrict ourselves to connections of the following type.
Every principal connection $B$ on
$P$ gives rise to the connection $\G_B$ (\ref{69}) on $C$ such that
\[\G_B\circ B={\cal S}\circ J^1B,\]
\[\G_B{}^m_{\m\la}=\frac{1}{2} [c^m_{nl}k^n_\la
k^l_\m  +\dr_\m B^m_\la+\dr_\la B^m_\m -c^m_{nl}
(k^n_\m B^l_\la+k^n_\la B^l_\m)] -\{^\bt{}_{\m\la}\}(B^m_\bt-k^m_\bt).\]

For all these connections, the Hamiltonian forms
\ben &&H_B=p^{\m\la}_mdk^m_\m\w\om_\la-
p^{\m\la}_m\G_B{}^m_{\m\la}\om-\wt{\cH}_{YM}\om, \label{5.3}\\
&&\wt{\cH}_{YM}= \frac{\ve^2}{4}a^{mn}_Gg_{\m\n}
g_{\la\bt} p^{[\m\la]}_m p^{[\n\bt]}_n| g| ^{-1/2},\nonumber\een
are associated with the Lagrangian density $L_{YM}$ and constitute a
complete family.
The corresponding Hamilton equations for sections $r$ of $\Pi\to X$ read
\ben &&\dr_\la p^{\m\la}_m=-c^n_{lm}k^l_\n
p^{[\m\n]}_n+c^n_{ml}B^l_\n p^{(\m\n)}_n
-\{^\m{}_{\la\n}\}p^{(\la\n)}_m, \label{5.5} \\
&&\dr_\la k^m_\m+ \dr_\m k^m_\la=2\G_B{}^m_{(\m\la)}\label{5.6}\een
plus the equation (\ref{5.2b}). The
equations (\ref{5.2b}) and (\ref{5.5}) restricted to the constraint space
(\ref{5.2a}) are the familiar Yang-Mills equations.
Different Hamiltonian forms (\ref{5.3}) lead to the different
equations (\ref{5.6}). The equation (\ref{5.6}) is independent of
canonical momenta and plays the role of the gauge-type condition.
Its solution is $k(x)=B$.

Turn now to the energy-momentum conservation law.

Let $A$ be a solution of the Yang-Mills equations. There exists the
Hamiltonian form $H_{B=A}$ (\ref{5.3}) such that $r_A=\wh L_{YM}\circ A$
is a solution of the corresponding Hamilton equations (\ref{5.5}),
(\ref{5.6}) and (\ref{5.2b}) on the constraint space (\ref{5.2a}).

Let us examine the conservation law (\ref{5.29}) where we take
$\wt{\cH}_\G=\wt{\cH}_{YM}$ and $\G=\G_{B=A}$ from the splitting (\ref{5.3}).

On the solution $r_A$, the curvature of the connection $\G_A$ is reduced to
\be
&& R^m_{\la\al\m}=\frac12(\dr_\la F^m_{\al\m}-
c^m_{qn}k^q_\la F^n_{\al\m}-\{^\bt{}_{\al\la}\} F^m_{\bt\m}-
\{^\bt{}_{\m\la}\}F^m_{\al\bt})=\\
&& \quad \frac12[(\dr_\al F^m_{\la\m} - c^m_{qn}k^q_\al F^n_{\la\m} -
\{^\bt{}_{\la\al}\}F^m_{\m\bt}) -
(\dr_\m F^m_{\la\al} - c^m_{qn}k^q_\m F^n_{\la\al} -
\{^\bt{}_{\la\m}\}F^m_{\al\bt})]
\ee
where $F=\cF\circ A$ is the strength of $A$. Set
\[S^\la{}_\m=p_m^{[\al\la]}\dr^m_{\al\m}\wt{\cH}_{YM}=
\frac{\ve^2}{2\sqrt{|g|}}a^{mn}_Gg_{\m\n}
g_{\al\bt} p^{[\al\la]}_m p^{[\bt\n]}_n.\]
We have
\[S^\la{}_\m = \frac12 p^{[\al\la]}\cF^m_{\m\al}, \qquad \wt{\cH}_{YM} =
\frac12 S^\al{}_\al.\]
In virtue of Eqs.(\ref{5.2a}), (\ref{5.2b}) and (\ref{5.5}), we
obtain the relations
\[\dr^\bt_n\G_A{}^m_{\al\m}p^{\al\la}_m\dr^n_{\bt\la}
\wt{\cH}_{YM}=\{^\bt{}_{\al\m}\}S^\al{}_\bt,\]
\[r_A{}_m^{[\la\al]}R^m_{\la\al\m}=\dr_\la S^\la{}_\m (r_A)
-\{^\bt{}_{\m\la}\}S^\la{}_\bt (r_A)\]
and find that
\[t^\la{}_\m\sqrt{|g|}=2S^\la{}_\m-\frac12\dl^\la_\m S^\al{}_\al,\]
\[T_{\G_A}{}^\la{}_\m(r_A)
=S^\la{}_\m(r_A)-\frac12\dl^\la_\m S^\al{}_\al(r_A),\]
\[t^\la{}_\m(r_A)\sqrt{|g|}=T_{\G_A}^\la{}_\m(r_A)+S^\la{}_\m(r_A).\]
Hence, the identity (\ref{5.29}) in gauge theory
is brought into the covariant energy-momentum conservation law
\[\nabla_\la t^\la{}_\m(r_A)\ap 0.\]

The Lagrangian partner of the Hamiltonian SEM tensor $T_{\G_A}(r_A)$ is
the SEM tensor $\cT_{\G_A}(A)$  (\ref{S14}) on the solution $A$
relative to the connection $\G_A$ on the
bundle $C$. This is exactly the familiar symmetrized canonical energy-momentum
tensor of gauge potentials.

Thus, we conclude that every solution in gauge
theory and, e.g., in electromagnetic theory requires
the own SEM tensor, otherwise in regular Lagrangian models.

\section{SEM tensors of matter fields}

In gauge theory, matter fields possessing only internal symmetries are
described by sections of a vector bundle
\[Y=(P\times V)/G\]
associated with a principal bundle $P$. It is provided with a $G$-invariant
fibre metric $a^Y$. Because of the canonical vertical splitting
$VY=Y\times Y$, the metric $a^Y$ is a
fibre metric in the vertical tangent bundle $VY\to X$.
Every principal connection $A$ on a principal bundle $P$ yields the
associated connection
\begin{equation}
\G=dx^\la\ot [\dr_\la +A^m_\m (x)I_m{}^i{}_jy^j\dr_i] \label{S4}
\end{equation}
where $A^m_\m (x)$ are coefficients of the local connection 1-form
and $I_m$ are generators of the structure group $G$
on the standard fibre $V$ of the bundle $Y$.

On the configuration space $J^1Y$, the regular Lagrangian density of
matter fields in the presence of a background connection $\G$ on $Y$ reads
\begin{equation}
L_{(m)}=\frac12a^Y_{ij}[g^{\m\n}(y^i_\m-\G^i_\m)
(y^j_\n-\G^j_\n)-m^2y^iy^j]\sqrt{| g|}\om.\label{5.12}
\end{equation}

The Legendre bundle of the vector bundle $Y$ is
\[\Pi=\op\wedge^n T^*X\op\ot_YTX\op\ot_Y Y^*. \]
The unique Hamiltonian form on $\Pi$ associated with the
Lagrangian density  $L_{(m)}$ (\ref{5.12}) is written
\begin{equation}
H_{(m)}=p^\la_idy^i\w\om_\la-p^\la_i
\G^i_\la\om- \frac12(a^{ij}_Yg_{\m\n}p^\m_ip^\n_j| g|^{-1}
+ m^2a^Y_{ij}y^iy^j)\sqrt{| g|}\om \label{5.13}
\end{equation}
where $a_Y$ is the fibre metric in $V^*Y$ dual to $a^Y$. There is the 1:1
correspondence between the solutions of the first order Euler-Lagrange
equations of the regular Lagrangian density (\ref{5.12})
and the solutions of the
Hamilton equations of the Hamiltonian form (\ref{5.13}).

To examine the conservation law (\ref{5.29}), let us take the same
Hamiltonian SEM tensor relative to the connection $\G$ (\ref{S4})
for all solutions $r$ of the Hamilton equations.
The following equality motivates the option above. We have
\be && T_\G^\la{}_\m(r)=t^\la{}_\m(r)\sqrt{|g|}=
[a^{ij}_Yg_{\m\n}r^\la_ip^\n_j|g|^{-1} \\
&& \qquad -\frac12 \dl^\la_\m(a^{ij}_Yg_{\al\n}r^\al_ir^\n_j|g|^{-1}+
m^2a^Y_{ij}r^ir^j)]\sqrt{|g|}.\ee
The gauge invariance condition
\[I_m{}^j{}_ir_j^\la\dr_\la^i\wt{\cH}=0\]
also takes place. Then, it easily observed that the identity (\ref{5.29})
reduces to the familiar covariant energy-momentum conservation law
\[\sqrt{|g|}\nabla_\la t^\la{}_\m(r)\ap -r^\la_iF^m_{\la\m}I_m{}^i{}_jy^j.\]

\section{SEM tensors in affine-metric gravitation theory}

After testing above, we apply the Hamiltonian SEM tensor machinery to
gravitation theory.

In this Section, $X$ is a 4-dimensional world manifold which obeys the
well-known topological conditions in order that a gravitational field
exists on $X^4$.

The contemporary concept of  gravitation interaction is based on
the gauge gravitation theory with  two types of gravitational
fields. These are tetrad garvitational fields and Lorentz gauge potentials.
In absence of fermion matter, one can choose by
gravitational variables a pseudo-Riemannian metric $g$ on a world
manifold $X^4$ and a general linear connections $K$
on the tangent bundle of $X^4$. We
call them a world metric and a world connection respectively.
Here we are not concerned with
the matter interecting with a general linear connection, for it
is non-Lagrangian and hypothetical as a rule.

Let $LX\to X^4$ be the principal bundle of linear frames in the
tangent spaces to $X^4$. Its structure group is $GL^+(4,{\bf R})$.
The world connections are associated with the principal connections on the
principal bundle $LX\to X^4$.
 Hence, there is the 1:1 correspondence between the
world connections and the global sections of the quotient bundle
\begin{equation}
C=J^1LX/GL^+(4,{\bf R}). \label{251}
\end{equation}
We therefore can apply the standard procedure of the Hamiltonian
gauge theory in order to describe the configuration and phase spaces of
world connections \cite{sard,7sar}.

 There is the 1:1 correspondence between the world metrics $g$ on
$X^4$  and the global sections of the bundle $\Si$ of
pseudo-Riemannian bilinear
forms in tangent spaces to $X^4$. This bundle is
associated with the $GL_4$-principal bundle $LX$.
The 2-fold covering of the bundle $\Si$ is the quotient bundle $LX/SO(3,1)$.

The total configuration space of the
affine-metric gravitational variables is the product
\begin{equation}
J^1C\op\times_{X^4}J^1\Sigma. \label{N33}
\end{equation}
coordinatized by $(x^\m, g^{\al\bt}, k^\al{}_{\bt\m}, g^{\al\bt}{}_\la,
k^\al{}_{\bt\m\la})$.

Also the total phase space $\Pi$ of the affine-metric gravity
is the product of the Legendre bundles
over the above-mentioned bundles $C$ and $\Si$.
It is provided with the corresponding canonical coordinates
$(x^\m, g^{\al\bt}, k^\al{}_{\bt\m},
p_{\al\bt}{}^\la, p_\al{}^{\bt\m\la})$.

On the configuration space (\ref{N33}), the  Hilbert-Einstein  Lagrangian
density of General Relativity reads
\begin{equation}
L_{HE}=-\frac{1}{2\kp}g^{\bt\la}\cF^\al{}_{\bt\al\la}
\sqrt{-g}\omega,\label{5.17}
\end{equation}
\[\cF^\al{}_{\bt\n\la}=k^\al{}_{\bt\la\n}-
k^\al{}_{\bt\n\la}+k^\al{}_{\ve\n}k^\ve{}_{\bt\la}-k^\al{}_{\ve\la}
k^\ve{}_{\bt\n}.\]
The corresponding Legendre morphism is given by the expressions
\ben
&& p_{\al\bt}{}^\la\circ \wh L_{HE}=0,\nonumber
\\ &&   p_\al{}^{\bt\n\la}\circ \wh L_{HE}
=\pi_\al{}^{\bt\n\la} =\frac{1}{2\kp}(\dl^\n_\al
g^{\bt\la}-\dl^\la_\al g^{\bt\n})\sqrt{-g}.\label{5.18}
\een
They define the constraint space of General Relativity in
the affine-metric variables.

Let us consider the following connections on the bundle $C\times\Si$
 in order to construct a complete family of Hamiltonian
forms associated with the Lagrangian density (\ref{5.17}).

Let $K$ be a world connection and
\be
&& \G_K{}^\al{}_{\bt\n\la}=\frac12 [k^\al{}_{\ve\n}
k^\ve{}_{\bt\la}-k^\al{}_{\ve\la}
k^\ve{}_{\bt\n} +\dr_\la K^\al{}_{\bt\n}+\dr_\n K^\al{}_{\bt\la}\\
&& \qquad -2K^\ve{}_{(\n\la)}(K^\al{}_{\bt\ve} -k^\al{}_{\bt\ve}) +
K^\ve{}_{\bt\la}k^\al{}_{\ve\n}+K^\ve{}_{\bt\n}k^\al{}_{\ve\la} -
K^\al{}_{\ve\la}k^\ve{}_{\bt\n}-K^\al{}_{\ve\n}k^\ve{}_{\bt\la}]
\ee
the corresponding connection on the bundle $C$ (\ref{251}). Let $K'$ be
another symmetric world connection. Building on these connections, we
set up the connection
\ben
&& \G^{\al\bt}{}_\la =-{K'}^\al{}_{\ve\la}g^{\ve\bt} -
{K'}^\bt{}_{\ve\la} g^{\al\ve}, \nonumber\\
&&\G^\al{}_{\bt\n\la} = \G_K{}^\al{}_{\bt\n\la}
-\frac12R^\al{}_{\bt\n\la} \label{N34}
\een
on the bundle $C\times\Si$ where $R^\al{}_{\bt\n\la}$ is the curvature of $K$.

For all connections (\ref{N34}), the Hamiltonian forms
\ben
&& H_{HE}=(p_{\al\bt}{}^\la dg^{\al\bt} +
p_\al{}^{\bt\n\la}dk^\al{}_{\bt\n})\w\om_\la
-\cH_{HE}\omega, \nonumber \\
&& \cH_{HE}=-p_{\al\bt}{}^\la({K'}^\al{}_{\ve\la}
g^{\ve\bt} +{K'}^\bt{}_{\ve\la} g^{\al\ve})
+p_\al{}^{\bt\n\la}\G_K{}^\al{}_{\bt\n\la} -\frac12R^\al{}_{\bt\n\la}
(p_\al{}^{\bt\n\la}-\pi_\al{}^{\bt\n\la})\nonumber\\
&&\qquad = -p_{\al\bt}{}^\la({K'}^\al{}_{\ve\la}
g^{\ve\bt} +{K'}^\bt{}_{\ve\la}g^{\al\ve})
+p_\al{}^{\bt\n\la}\G{}^\al{}_{\bt\n\la}+\wt{\cH}_{HE},  \label{5.19}\\
&&\wt{\cH}_{HE} = \frac1{2\kp}R\sqrt{-g},
\een
are associated with the Lagrangian density $L_{HE}$ and constitute a
complete family.

Given the Hamiltonian form $H_{HE}$ (\ref{5.19}) plus
a Hamiltonian form $H_M$ for matter,
the corresponding Hamilton equations read
\bea
&&\dr_\la g^{\al\bt} +{K'}^\al{}_{\ve\la}g^{\ve\bt}
+{K'}^\bt{}_{\ve\la}g^{\al\ve}=0, \label{5.20a}\\
&&\dr_\la k^\al{}_{\bt\n}=\G_K{}^\al{}_{\bt\n\la}
-\frac12R^\al{}_{\bt\n\la}, \label{5.20b} \\
&&\dr_\la p_{\al\bt}{}^\la =p_{\ve\bt}{}^\si
{K'}^\ve{}_{\al\si} +p_{\ve\al}{}^\si {K'}^\ve{}_{\bt\si}
-\frac{1}{2\kp}(R_{\al\bt} -\frac12g_{\al\bt}R)\sqrt{-g}
-\frac{\dr\cH_M}{\dr g^{\al\bt}},\label{5.20c} \\
&& \dr_\la p_\al{}^{\bt\n\la}= -p_\al{}^{\ve[\n\g]}
k^\bt{}_{\ve\g} +p_\ve{}^{\bt[\n\g]}
k^\ve{}_{\al\g} - p_\al{}^{\bt\ve\g}K^\n{}_{(\ve\g)} \nonumber\\
&&\qquad -p_\al{}^{\ve(\n\g)}K^\bt{}_{\ve\g} +p_\ve{}^{\bt(\n\g)}
K^\ve{}_{\al\g} \label{5.20d}
\eea
plus the motion equations of matter.
The Hamilton equations (\ref{5.20a}) and (\ref{5.20b}) are independent
of canonical momenta and so, reduce to the gauge-type conditions.
The equation (\ref{5.20b}) breaks into the following two parts
\ben
&& F^\al{}_{\bt\la\n}=R^\al{}_{\bt\n\la},\label{5.21}\\
&& \dr_\n(K^\al{}_{\bt\la} -k^\al{}_{\bt\la})
+\dr_\la(K^\al{}_{\bt\n} -k^\al{}_{\bt\n}) -2K^\ve{}_{(\n\la)}
(K^\al{}_{\bt\ve} -k^\al{}_{\bt\ve}) \nonumber\\
&& \qquad +K^\ve{}_{\bt\la}k^\al{}_{\ve\n}
+K^\ve{}_{\bt\n}k^\al{}_{\ve\la} -K^\al{}_{\ve\la}k^\ve{}_{\bt\n}
-K^\al{}_{\ve\n}k^\ve{}_{\bt\la}=0 \label{5.22}
\een
where $F$ is the curvature of the connection $k(x)$. It is readily observed
that the gauge-type conditions (\ref{5.20a}) and (\ref{5.20b})
are satisfied by
\begin{equation}
k(x)=K, \qquad {K'}^\al{}_{\bt\la}= \{^\al{}_{\bt\la}\}.\label{J13}
\end{equation}

When restricted to the constraint space (\ref{5.18}), the Hamilton
equations (\ref{5.20c}) and (\ref{5.20d}) come to
\ben
&& \frac{1}{\kp}(R_{\al\bt}
-\frac12 g_{\al\bt}R)\sqrt{-g}= -\frac{\dr\cH_M}{\dr g^{\al\bt}},
\label{5.23} \\
&& D_\al(\sqrt{-g}g^{\n\bt}) - \dl^\n_\al
D_\la(\sqrt{-g}g^{\la\bt}) +\sqrt{-g}[g^{\n\bt}(k^\la{}_{\al\la} -
k^\la{}_{\la\al}) \nonumber\\
&& \qquad + g^{\la\bt}(k^\n{}_{\la\al}-k^\n{}_{\al\la})+ \dl^\n_\al
g^{\la\bt} (k^\m{}_{\m\la} - k^\m{}_{\la\m})] =0 \label{5.24}
\een
where
\[D_\la g^{\al\bt}= \dr_\la g^{\al\bt} + k^\al{}_{\m\la}g^{\m\bt} +
k^\bt{}_{\m\la}g^{\al\m}.\]
Substituting the equation (\ref{5.21}) into the equation (\ref{5.23}),
we obtain the Einstein equations
\begin{equation}
\frac{1}{\kp}(F_{\al\bt}-\frac12 g_{\al\bt}F)= -t_{\al\bt}\label{5.25}
\end{equation}
where $t_{\al\bt}$ is the metric energy-momentum tensor of matter.
The equations(\ref{5.24}) and (\ref{5.25}) are the
familiar equations of affine-metric gravity. In particular,
the former is the equation for
torsion and nonmetricity terms of the general linear connection $k(x)$.
In the absence of matter sources of a general linear connection,
it admits the well-known solution
\be
&&k^\al{}_{\bt\n} =\{^\al{}_{\bt\n}\} - \frac12\dl^\al_\n V_\bt,\\
&& D_\al g^{\bt\g}= V_\al g^{\bt\g},
\ee
where $V_\al$ is an arbitrary covector field corresponding to the
well-known projective freedom.

Turn now to the identity (\ref{5.27}).

Let $s=(k(x), g(x))$ be a
solution of the Euler-Lagrange equations of the first order Hilbert-Einstein
Lagrangian density (\ref{5.17}) and $r$ the corresponding solution of
the Hamilton equations of the Hamiltonian form (\ref{5.19})
where $K$ and $K'$ are given by the expressions (\ref{J13}). For this
solution $r$, let us take the SEM tensor $T_s$ (\ref{J3})
relative to the connection (\ref{N34}) where
$K$ and $K'$ are given by the expressions (\ref{J13}). It reads
\[T_s{}^\la{}_\m = \dl^\la_\m \wt{\cH}_{HE}=
\frac{1}{2\kp}\dl^\la_\m R\sqrt{-g}\]
and the identity (\ref{5.27}) takes the form
\ben
&&(\dr_\m+\G^{\al\bt}{}_\m\dr_{\al\bt}+\G^i_\m\dr_i-\dr_i\G^j_\m
p^\la_j\dr^i_ \la)( \wt{\cH}_{HE}+\wt{\cH}_M) \ap \nonumber\\
&& \qquad \frac{d}{dx^\la}(
T_s{}^\la{}_\m+T_M{}^\la{}_\m)+ p_\al{}^{\bt\n\la}R^\al{}_{\bt\n\la\m}
 + p^\la_iR^i_{\la\m} \label{E6}
\een
where $T_M$ is the SEM tensor for matter.

One can verify that the SEM tensor $T_s$ meets the condition
\begin{equation}
(\dr_\m +\G^{\al\bt}{}_\m\dr_{\al\bt})\wt\cH_{HE}=
\frac{d}{dx^\la}T_s{}^\la{}_\m \label{J14}
\end{equation}
and, on solutions (\ref{J13}), the curvature
of the connection (\ref{N34}) vanishes. Hence, the identity (\ref{E6})
is reduced to the conservation law (\ref{5.29}) of matter in the
presence of a background metric.
The gravitation SEM tensor is eliminated from the conservation law
because the Hamiltonian form $\cH_{HE}$ is affine in all
canonical momenta. Note that only gauge-type conditions (\ref{5.20a}),
(\ref{5.20b}) and the motion equations of matter have been used.

At the same time, since the canonical momenta $p_{\al\bt}{}^\la$ of the
world metric are equal to zero, the Hamilton equation (\ref{5.20c}) on
the Lagrangian constraint space comes to
\[\dr_{\al\bt}( \wt{\cH}_{HE}+\wt{\cH}_M)=0.\]
Hence, the equality (\ref{E6}) takes the form
\begin{equation}
\pi_\al{}^{\bt\n\la}\dr_\m R^\al{}_{\bt\n\la} +
(\dr_\m+\G^i_\m\dr_i-\dr_i\G^j_\m p^\la_j\dr^i_ \la)\wt{\cH}_M\ap
\frac{d}{dx^\la}(T_s{}^\la{}_\m+T_M{}^\la{}_\m)+ p^\la_iR^i_{\la\m}.\label{E7}
\end{equation}
It is the form of the energy-momentum conservation law which
we observe also in case of quadratic Lagrangian densities of
affine-metric gravity. Substituting the equality (\ref{J14}) into
(\ref{E7}), we obtain the above mentioned result.

As a test case of quadratic Lagrangian densities of affine-metric gravity,
let us examine the sum
\begin{equation}
L = (-\frac{1}{2\kp}g^{\bt\la}\cF^\al{}_{\bt\al\la} +\frac{1}{4\ve}
g_{\al\g}g^{\bt\si}g^{\n\m}g^{\la\ve}\cF^\al{}_{\bt\n\la}
\cF^\g{}_{\si\m\ve})\sqrt{-g}\om \label{E18}
\end{equation}
of the Hilbert-Einstein Lagrangian density and the Yang-Mills one.
The corresponding Legendre morphism reads
\bea
&& p_{\al\bt}{}^\la\circ\wh L=0,\label{E9a}\\
&& p_\al{}^{\bt(\n\la)}\circ\wh L=0, \label{E9b}\\
&& p_\al{}^{\bt[\n\la]}\circ\wh L=\pi_\al{}^{\bt\n\la} +\frac1\ve
g_{\al\g}g^{\bt\si}g^{\n\m}g^{\la\ve}\cF^\g{}_{\si\ve\m}\sqrt{-g}.
\label{E9c}
\eea
The relations (\ref{E9a}) and (\ref{E9b}) defines the Lagrangian
constraint space.

Let us consider connections
\ben
&& \G^{\al\bt}{}_\la =-{K'}^\al{}_{\ve\la}g^{\ve\bt} -
{K'}^\bt{}_{\ve\la} g^{\al\ve}, \nonumber\\
&&\G^\al{}_{\bt\n\la} = \G_K{}^\al{}_{\bt\n\la} \label{E10}
\een
on the bundle $C\times\Si$ where the notations of the expression
(\ref{N34}) are utilized. The corresponding Hamiltonian forms
\ben
&& H=(p_{\al\bt}{}^\la dg^{\al\bt} +
p_\al{}^{\bt\n\la}dk^\al{}_{\bt\n})\w\om_\la-\cH\om, \nonumber \\
&& \cH=-p_{\al\bt}{}^\la({K'}^\al{}_{\ve\la}g^{\ve\bt} +{K'}^\bt{}_{\ve\la}
g^{\al\ve})+p_\al{}^{\bt\n\la}\G_K{}^\al{}_{\bt\n\la}+\wt{\cH},\nonumber \\
&& \wt{\cH}=\frac{\ve}4 g^{\al\g}g_{\bt\si}g_{\n\m}g_{\la\ve}
(p_\al{}^{\bt[\n\la]}-\pi_\al{}^{\bt\n\la}) (p_\g{}^{\si[\m\ve]}
-\pi_\g{}^{\bt\m\ve}),  \label{E11}
\een
are associated with the Lagrangian density ({\ref{E18}) and constitute
a complete family.

Given the Hamiltonian form (\ref{E11}) plus the Hamiltonian form $H_M$
for matter, the corresponding Hamilton equations read
\bea
&&\dr_\la g^{\al\bt} +{K'}^\al{}_{\ve\la}g^{\ve\bt}
+{K'}^\bt{}_{\ve\la}g^{\al\ve}=0, \label{E12a}\\
&&\dr_\la k^\al{}_{\bt\n}=\G_K^\al{}_{\bt\n\la} +
\ve g^{\al\g}g_{\bt\si}g_{\n\m}g_{\la\ve}(p_\g{}^{\si[\m\ve]}
-\pi_\g{}^{\bt\m\ve}) , \label{E12b} \\
&&\dr_\la p_{\al\bt}{}^\la =-\frac{\dr\cH}{\dr g^{\al\bt}}
-\frac{\dr\cH_M}{\dr g^{\al\bt}},\label{E12c} \\
&& \dr_\la p_\al{}^{\bt\n\la}= -p_\al{}^{\ve[\n\g]}
k^\bt{}_{\ve\g} +p_\ve{}^{\bt[\n\g]}k^\ve{}_{\al\g} \nonumber \\
&& \qquad - p_\al{}^{\bt\ve\g}K^\n{}_{(\ve\g)} -p_\al{}^{\ve(\n\g)}
K^\bt{}_{\ve\g} +p_\ve{}^{\bt(\n\g)}K^\ve{}_{\al\g} \label{E12d}
\eea
plus the motion equations for matter.
The equation (\ref{E12b}) breaks into the equation (\ref{E9c}) and the
gauge-type condition (\ref{5.22}).
The gauge-type conditions (\ref{E12a}) and (\ref{5.22}) have the
solution (\ref{J13}).
Substituting the equation (\ref{E12b}) into the equation
(\ref{E12c}) on the constraint space (\ref{E9a}), we get the quadratic
Einstein equations. Substitution of the equations
(\ref{E9b}) and (\ref{E9c}) into the equation (\ref{E12d}) results into
the Yang-Mills generalization
\[\dr_\la p_\al{}^{\bt\n\la} +p_\al{}^{\ve[\n\g]}
k^\bt{}_{\ve\g} -p_\ve{}^{\bt[\n\g]}k^\ve{}_{\al\g} =0\]
of the equation (\ref{5.24}).

Turn now to the energy-momentum conservation law.
Let us consider the splitting of the Hamiltonian form
(\ref{E11}) with respect to the connection (\ref{N34}) and the
Hamiltonian density
\[\wt{\cH}_\G= \wt{\cH} +\frac12 p_\al{}^{\bt\n\la}R^\al{}_{\bt\n\la}.\]

Let $s=(k(x), g(x))$ be a solution of the Euler-Lagrange equations of the
Lagrangian density (\ref{E18}) and $r$ the corresponding solution of
the Hamilton equations of the Hamiltonian form (\ref{E11})
where $K$ and $K'$ are given by the expressions (\ref{J13}). For this
solution $r$, let us take the SEM tensor $T_s$ (\ref{J3})
relative to the connection (\ref{N34}) where
$K$ and $K'$ are given by the expressions (\ref{J13}). It reads
\be
&& T_s{}^\la{}_\m = \frac12 p_\al{}^{\bt[\n\la]}R^\al{}_{\bt\n\m} +
\frac{\ve}{2}g^{\al\g}g_{\bt\si}g_{\n\dl}g_{\m\ve}
p_\al{}^{\bt[\n\la]} (p_\g{}^{\si[\dl\ve]}-\pi_\g{}^{\si\dl\ve})\\
&&\qquad -\dl^\la_\m (\wt{\cH} +
\frac{\ve}{2}g^{\al\g}g_{\bt\si}g_{\n\dl}g_{\tau\ve}
\pi_\al{}^{\bt\n\tau} (p_\g{}^{\si[\dl\ve]}-\pi_\g{}^{\si\dl\ve}))
\ee
and is equal to
\[\frac1\ve R_\al{}^{\bt\n\la}R^\al{}_{\bt\n\m}
+ \pi_\al{}^{\bt\n\la}R^\al{}_{\bt\n\m}
-\dl^\la_\m (\frac{1}{4\ve} R_\al{}^{\bt\n\la}R^\al{}_{\bt\n\la}+
\frac{1}{2\kp}R).\]
The identity (\ref{5.27}) takes the form
\be
&&(\dr_\m+\G^{\al\bt}{}_\m\dr_{\al\bt}+\G^i_\m\dr_i-\dr_i\G^j_\m
p^\la_j\dr^i_ \la-p_\al{}^{\bt\n\la} \frac{\dr}{\dr k^\si{}_{\g\dl}}
\G_K{}^\al{}_{\bt\n\m}\frac{\dr}{\dr P_\si{}^{\g\dl\la}})
(\wt{\cH}_\G+\wt{\cH}_M) \ap \\
&& \qquad \frac{d}{dx^\la}(
T_s{}^\la{}_\m+T_M{}^\la{}_\m)+ p_\al{}^{\bt\n\la}R^\al{}_{\bt\n\la\m}
 + p^\la_iR^i_{\la\m}
\ee
and is simplified to
\begin{equation}
(\dr_\m+\G^i_\m\dr_i-\dr_i\G^j_\m p^\la_j\dr^i_ \la)\wt{\cH}_M
-p_\al{}^{\bt\n\la} \frac{\dr}{\dr k^\si{}_{\g\dl}}
\G_K{}^\al{}_{\bt\n\m}\frac{\dr}{\dr p_\si{}^{\g\dl\la}}
\wt{\cH}_\G\ap
\frac{d}{dx^\la}(T_s{}^\la{}_\m+T_M{}^\la{}_\m)+ p^\la_iR^i_{\la\m}
\label{E20}
\end{equation}
where
\begin{equation}
p_\al{}^{\bt\n\la} \frac{\dr}{\dr k^\si{}_{\g\dl}}
\G_K{}^\al{}_{\bt\n\m}\frac{\dr}{\dr p_\si{}^{\g\dl\la}}
\wt{\cH}_\G= \frac1\kp k^\g{}_{\bt\m}(g^{\bt\n}R^\al{}_{\g\al\n} -
g^{\al\n}R^\bt{}_{\al\n\g})\sqrt{-g} -
k^\g{}_{(\bt\m)}p_\al{}^{\n\bt\la}R^\al{}_{\n\g\la}.\label{J15}
\end{equation}
Note that if $k{x}$ is a Lorentz connection, the term (\ref{J15}) comes to
\[-k^\g{}_{(\bt\m)}p_\al{}^{\n\bt\la}R^\al{}_{\n\g\la}.\]

Let us choose the local geodetic coordinate system at a point $x\in X$.
Relative to this coordinate system, the equality (\ref{E20}) at $x$
comes to the conservation law
\[(\dr_\m+\G^i_\m\dr_i-\dr_i\G^j_\m p^\la_j\dr^i_ \la)\wt{\cH}_M
\ap \frac{d}{dx^\la}(T_s{}^\la{}_\m+T_M{}^\la{}_\m)+ p^\la_iR^i_{\la\m}.\]
For instance, in gauge theory, we have
\[\frac{d}{dx^\la}(T_\G{}^\la{}_\m+t_M{}^\la{}_\m)=0\]
where $t_M$ is the metric energy-momentum tensor of matter.
\bigskip

\centerline{\large\bf Acknowledgement}
\medskip

The author thanks L.Mangiarotti and G.Giachetta (University of Camerino)
for fruitful discussion.

\end{document}